\documentstyle[12pt]{article}

\newcommand \beq{\begin{eqnarray}}
\newcommand \eeq{\end{eqnarray}}
\def\del{\partial}
\def\frac#1#2{{#1 \over #2}}

\def\half{\ifinner {\scriptstyle {1 \over 2}}
   \else {1 \over 2} \fi}

\def\simge{\mathrel{%
   \rlap{\raise 0.511ex \hbox{$>$}}{\lower 0.511ex \hbox{$\sim$}}}}
\def\simle{\mathrel{
   \rlap{\raise 0.511ex \hbox{$<$}}{\lower 0.511ex \hbox{$\sim$}}}}

\def\slashchar#1{\setbox0=\hbox{$#1$}
   \dimen0=\wd0
   \setbox1=\hbox{/} \dimen1=\wd1
   \ifdim\dimen0>\dimen1
      \rlap{\hbox to \dimen0{\hfil/\hfil}}
      #1
   \else
      \rlap{\hbox to \dimen1{\hfil$#1$\hfil}}
      /
   \fi}

\def\journal#1#2#3#4{\ {#1}{\bf #2} ({#3})\  {#4}}

\def\NPA{\journal{Nucl.\ Phys.\ {\bf A}}}
\def\NPB{\journal{Nucl.\ Phys.\ {\bf B}}}

\def\PLB{\journal{Phys.\ Lett.\ {\bf B}}}

\def\PRD{\journal{Phys.\ Rev.\ {\bf D}}}
\def\PRL{\journal{Phys.\ Rev.\ Lett.}}
\def\PhysRept{\journal{Phys.\ Repts.}}

\def\SovPhysJETP{\journal{Sov.\ Phys.\ JETP}}

\def\SovJNuclPhys{\journal{Sov.\ J.\ Nucl.\ Phys.}}

\begin{document}
\begin{titlepage}
\begin{center}
{\Large KINETIC EQUATIONS FOR LONGWAVELENGTH\\ EXCITATIONS\\
OF THE QUARK-GLUON PLASMA\\}
\vskip 1cm
Jean-Paul BLAIZOT\footnote{CNRS}  and
Edmond IANCU\\
Service de Physique Th\'eorique\footnote{Laboratoire de la Direction des
Sciences de la Mati\`ere du Commissariat \`a l'Energie
Atomique}, CE-Saclay \\ 91191 Gif-sur-Yvette, France\\
\vskip0.5cm
January 8, 1993
\end{center}
\vskip 2.5cm \begin{abstract} We show that longwavelength excitations of
the quark-gluon plasma are described by simple kinetic equations which
represent
 the exact equations of motion at leading order in $g$. Properties of the
so-called ``hard
thermal loops'', i.e.  the dominant contributions to amplitudes
with soft external lines, find in this approach
a natural explanation. In particular, their generating functional appears here
as the effective action describing long wavelength excitations of the plasma.
\end{abstract}
\vskip 4cm
\begin{flushright}
SPhT/93-1
\end{flushright}
\begin{flushleft}
Submitted to Physical Review Letters\\
PACS No: 12.38.Mh, 12.38.Bx, 52.25.Dg, 11.15.Kc
\end{flushleft}
\end{titlepage}
\setcounter{equation}{0}
Significant progress has been achieved recently in
understanding long  wavelength
excitations of a  quark-gluon plasma\cite{Pisarski89,Braaten90,Baym90}.
 In  equilibrum,  at high temperature, such a  plasma may be viewed as a
gas of weakly interacting,
 massless quarks and gluons. When coupled to weak and slowly varying
perturbations, this system may acquire a collective behaviour on a length scale
$\sim 1/gT$, where $T$ is the temperature and $g$ the coupling constant,
assumed
to be small. In this letter, we present a consistent and physically intuitive
description of such long wavelength phenomena, based on  a set of coupled mean
field and kinetic
 equations, to  which
 the exact equations of motion
 reduce in leading order in $g$\cite{QED,QCD}.

The kinetic equations encompass all the so-called ``hard thermal
loops'' and clarify the nature of their
remarkable properties, left largely unexplained by their
original derivation in  terms of Feynman diagrams\cite
{Shuryak78,Frenkel90,Braaten90}.
These hard thermal
loops (HTL)
are the dominant corrections, at high temperature, to amplitudes involving
soft external lines. (Following the
usual terminology, we call  an energy or a momentum ``soft'' when it is of
order $gT$, and ``hard'' when it is of order $T$; at equilibrum, most
particles are hard.) As they are of the same
order of magnitude as the corresponding  non-vanishing tree level amplitudes,
HTL need to be resummed consistently in higher order
calculations\cite{Pisarski89,Braaten90}.

Our equations isolate consistently the dominant terms in $g$
 in the hierarchy of equations
which describe the response of the plasma to weak and slowly varying
disturbances, i.e. varying on a scale of order $1/gT$. In doing so, we treat
bosons and fermions on the same footing and  introduce an
average fermionic field $\psi(X)$ in parallel with the average gauge field
$A_\mu(X)$. A noteworthy feature of the present problem is that $g$, besides
measuring the interaction strength,  controls  the wavelength of the soft
space-time variations and, for consistency, also the strength of the mean
fields. One finds for example that, in order for the deviations away from
equilibrium to stay small,  the gauge field strength tensor should be at most
of order $gT^2$; then $gA\sim gT$ is of the same order as the derivative of a
``slowly varying'' quantity\cite{QED,QCD}.

  The  dominant
interactions which determine  the response of the plasma  are those which take
place between the hard particles and  the soft mean fields.  In leading order,
we
can neglect  the direct interaction between the hard particles, which allows us
to truncate the equations at the level of the 2-point functions. The motion of
a
given hard particle is only slightly perturbed by its interaction with a soft
mean field.
 However, because the mean fields vary over distances much larger than
the interparticle distance $(\sim 1/T)$, they affect coherently many hard
particles, giving rise to collective ``polarization''
phenomena. These  show up as ``induced sources'' which add to the external ones
in determining the properties of the mean fields.

The equations for the quark and gauge average fields  are
\beq
\label{4avpsi}
i\slashchar{D}_X \psi(X)=\eta(X)+\eta^{ind}(X),
\eeq
and
\beq
\label{4avA}
\left [\, D_X^\nu,\, F_{\nu\mu}(X)\,\right ]^a
-g  \bar\psi (X)\gamma_\mu t^a \psi(X)
=\,j_\mu^a(X)+j_\mu^{ind\, a}(X).
\eeq
Here $\eta$ and $j_\mu^a$ are external sources, $a=1,\ldots,N^2-1$ are
color indices for the adjoint representation
of the $SU(N)$ gauge group, while $\mu,\,\nu = 0,\ldots,d-1$ are space-time
indices ($d=4$ throughout this work). The covariant derivative is $D_\mu \equiv
\del_\mu+igt^a A_\mu^a$ and $F_{\mu\nu}\equiv [D_\mu,D_\nu]/(ig)$.
When using a covariant gauge, one should also consider equations for ghost
mean fields. These are not written here as it turns out that they are trivial,
i.e there are no induced sources for the ghost mean fields\cite{QCD}. After
functional
differentiation, the induced sources yield  the one particle irreducible
amplitudes with soft external lines. Thus for
 example the fermion self-energy is given by
$\Sigma(x,y)=\delta\eta^{ind}(x)/\delta\psi(y)$, the gluon polarization tensor
by $\Pi^{ab}_{\mu\nu}(x,y)=\delta j^{ind\, a}_\mu (x)/\delta  A_b^\nu(y)$,
etc...

In leading order, the induced sources can be expressed entirely
in terms of 2-point functions. For example,
$\eta^{ind}(x)=g\gamma^\nu t_a\langle A_\nu^a (x)\psi(x)\rangle_c$,
where the subscript $c$ indicates a connected expectation value which, in
leading order, involves only the hard particles. The abnormal
quark-gluon propagator which enters $\eta^{ind}$ is
 nonvanishing only in the presence of the  fermionic mean field $\psi$.
It vanishes in equilibrum, as does the induced
color current $j^{ind}$. This current  receives contributions from both
fermionic and bosonic particles, and accordingly may be written as
$j^{ind}=j_{\rm f}+j_{\rm b}$. The quark contribution is $j^\mu_{{\rm f}\,
a}=g\langle\bar\psi(x)\gamma^\mu t_a\psi(x)\rangle_c$. In an arbitrary
covariant
gauge, the bosonic piece  involves  contributions from both  hard gluons and
hard ghosts \cite{QCD}.  The polarization phenomena may be induced either by a
gauge field $A$, or by a fermionic one, $\psi$. Correspondingly, we set
$j_{\rm f}=j^A_{\rm f}+ j^\psi_{\rm f}$ and $j_{\rm b}=j_{\rm b}^A+j_{\rm
b}^\psi$. Here, for instance, $j^A_{\rm f}$ is the color current associated
to the collective motion of hard fermions induced by
the soft mean field $A$. It is independent of the fermionic field $\psi$. In
contrast, we shall see that $j^\psi$ has  generally a dependence  on $A$
imposed
by gauge covariance. Note also that the ghosts do not contribute to
$j^\psi_{\rm
b}$, as they have no direct interaction with the fermionic mean field.

In order to implement the condition that the average fields are slowly
varying, it is convenient to use the Wigner transform
 of the 2-point functions, such as
\beq\label{calS}
{\cal S}(k,X)\equiv\int d^4 s e^{ik\cdot s}
\langle {\rm T}\psi (X+\frac {s}{2})\bar\psi(X-\frac{s}{2})\rangle_c,
\eeq
for the quark propagator.
Note that, in contrast to other authors, we do not insist on defining
manifestly
gauge covariant Wigner functions \cite{Heinz83}.
Covariance will be recovered later, when calculating physical quantities.
The induced sources are then integrals over $k$ of  Wigner functions,
which we refer to as $k$-space densities. For example:
 \beq\label{4eind}
\eta^{ind}(X)=g\int\frac{d^4k}{(2\pi)^4}t^a \gamma^\nu{\cal K}^a_\nu(k,X),
\eeq
where ${\cal K}^a_\nu(k,X)$ is the Wigner transform of
the quark-gluon propagator $K_\nu^a(x,y)\equiv\\ \langle{\rm T}\psi(x)A_
\nu^a(y)\rangle_c$.  Thus ${\slashchar {\cal
K}}(k,X)\equiv t^a {\slashchar {\cal
K}}^a(k,X)\equiv t^a\gamma^\nu {\cal K}^a_\nu(k,X)$ is the $k$-space density
for $\eta^{ind}(X)$. The densities for the induced currents  are denoted
by ${\cal J}_{{\rm f},\,{\rm b}}^A(k,X)$ and ${\cal J}_{{\rm f},\,{\rm b}}^
\psi(k,X)$. For example, the fermionic density is
${\cal J}_{{\rm f}\, \mu}^{\, a}(k,X)={\rm Tr}\gamma_\mu
t^a {\cal S}(k,X)$, where the trace refers to both spin and color indices.

At leading order in the coupling $g$, the equations for
the densities of the induced sources read\cite{QCD}
\beq \label{4kinK} \left (k\cdot D_X\right ) {\cal {\slashchar
K}}(k,X) =-i\frac{g}{2}(d-2)C_f(\Delta(k)+\tilde\Delta(k))\slashchar{k}
\psi(X),
\eeq
\beq
\label{4Vlas}
 \left [k\cdot D_X,\, {\cal J}^A_{{\rm f}\, \mu}(k,X) \right ]^a=
2g\,N_f\,k_\mu\,k^\rho F_{\rho\nu}^a\del^\nu\tilde\Delta(k),
\eeq
\beq
\label{4Spsi}
\,\qquad\left [k\cdot D_X,\,{\cal J}^\psi_{{\rm f}\, \mu}(k,X) \right
]^a&=&ig\,k_\mu
\,\left \{\bar\psi(X)\,t^a\,\slashchar{{\cal K}}(k,X)
-\slashchar{{\cal H}}(k,X)\,t^a\,\psi(X)\right \}\nonumber\\
&-&g\,k_\mu\,f^{abc}\,\left \{\bar\psi(X)\,t^b\,\slashchar{{\cal K}}^c(k,X)
-\slashchar{{\cal H}}^b(k,X)\,t^c\,\psi(X)\right \},\qquad\qquad
\eeq
\beq
\label{4kJgpsi}
\left [k\cdot\,D_X,\,{\cal J}^\psi_{{\rm b}\, \mu}(k,X)\right ]^a=
\,g\,k_\mu\,f^{abc}\,\left \{\bar\psi(X)\,t^b\,\slashchar{{\cal K}}^c(k,X)
-\slashchar{{\cal H}}^b(k,X)\,t^c\,\psi(X)\right \},
\eeq
\beq
\label{covJg}
 \left [k\cdot D_X,\,{\cal J}^A_{{\rm b}\,\,\mu}(k,X) \right ]^a=
g\,N(d-2)\,k_\mu\,k^\rho F_{\rho\nu}^a\del^\nu\Delta(k).
\eeq
In these equations, $\slashchar{{\cal H}}(k,X)=\slashchar{{\cal
K}}^\dagger(k,X)
\gamma^0$, $C_f\equiv (N^2-1)/2N$ is the quark Casimir, $N_f$ is
the number of quark flavors and
$f^{abc}$ denote the structure constants of $SU(N)$. Furthermore,
$\Delta(k)\equiv\rho_0(k)N(k_0)$ and $\tilde\Delta(k)\equiv\rho_0(k)n(k_0)$,
where
$\rho_0(k)=2\pi\epsilon(k_0)\delta(k^2)$ is the spectral function for free
massless particles, and $N(k_0)$ and $n(k_0)$ denote respectively  boson and
 fermion occupation factors.
The factor $(d-2)$ in Eqs.~(\ref{4kinK}) and (\ref{covJg}) reflects the fact
that
only the
 transverse gluons effectively contribute to the densities.

Eqs.~(\ref{4kinK}-\ref{covJg}) have a number of interesting
properties: {\bf i)} They are independent of the gauge fixing parameter
$\lambda$
which enters calculations in general covariant gauges \cite{QED,QCD}.
{\bf ii)} In their right hand sides, all possible vacuum contributions cancel.
{\bf iii)} They transform covariantly under a local gauge
transformation of  the mean fields $A_\mu$, $\psi$ and $\bar\psi$.
The densities
${\cal J}_{\rm f}^\psi$, ${\cal J}_{\rm b}^\psi$ and $\slashchar{{\cal K}}$
 involve Wigner transforms which are gauge covariant in leading order.
The current induced by a gauge field involves   $k$-space densities,  ${\cal
J}_{\rm f}^A$ and ${\cal J}_{\rm b}^A$, which are derived from non covariant
Wigner functions. However, these densities are defined up to a total
derivative with respect to $k$ which does not contribute to the integrated
current. We have used this freedom  in order to make the densities explicitly
covariant \cite{QED,QCD}. {\bf iv)} The symmetry between Eqs.~(\ref{4Vlas})
and (\ref{covJg}) reflects the fact that hard quarks and gluons respond
similarly to a soft gauge field.  The same symmetry is apparent in
Eq.~(\ref{4kinK})  expressing the effect of the fermionic mean field on the
hard particles. {\bf v)} Note finally the presence of the factor $\delta(k^2)$
 in the r.h.s of these equations. This reflects the elementary dynamics of
the hard particles: they remain on their unperturbed mass shell and
only undergo essentially forward scattering on the mean fields.

As these remarks strongly suggest, the motion of the hard particles described
by Eqs.~(\ref{4kinK}-\ref{covJg})
 exhibits many features of classical dynamics. This becomes
more transparent if one makes explicit the structure of the various densities
implied by these equations. For instance, Eq.~(\ref{4Vlas}) implies
  \beq
{\cal J}^A_{{\rm f}\,\,\mu}(k,X)=2\,k_\mu\,N_f\,t^a (2\pi \delta(k^2))
\left [\theta (k^0)\delta n^a_+({\vec k},X) + \theta (-k^0) \delta n^a_-
(-{\vec k},X)\right ],
\eeq
where $\delta n_\pm=\delta n^a_\pm t^a$ are fluctuations in the quark color
densities induced  by the gauge field. These fluctuations satisfy (with
$\epsilon_k\equiv |\vec k|$) \beq
\left[ v\cdot D_X,\delta n_\pm({\vec k},X)\right]^a=\mp g\vec
v\cdot\vec E^a(X)\frac{dn(\epsilon_k)}{d\epsilon_k}.
\eeq
In the abelian case, this equation coincides with the linearized Vlasov
equation. Here, the color electric field not only modifies the motion of the
particle, but also induces a ``precession'' of the densities in color space.
The other equations may be given similar interpretation. Thus,
 Eq.~(\ref{4kinK}) for the quark-gluon Wigner
function ${\slashchar {\cal K}}$  describes fluctuations where, under the
action of a soft fermionic mean field,
  quarks are converted into  gluons and  vice-versa.

The total current induced by a gauge field $A$ is $j^A=j^A_{\rm f}+
j^A_{\rm b}$. Its $k$-space density, ${\cal J}^A\equiv {\cal J}^A_{\rm f}+
{\cal J}^A_{\rm b}$, satisfies
\beq
\label{4JA}
\left [k\cdot D_X,\, {\cal J}^A_{\mu}(k,X) \right ]=
g\,\,k_\mu\,k\cdot F(X)\cdot \del\left ( 2 N_f\tilde\Delta(k)+
N(d-2)\Delta(k)\right ).
\eeq
This equation generalizes the Vlasov equation to nonabelian plasmas.
Previous attempts to derive such an equation led to more intricate
results. However, they were based on
different  approximation schemes which  mix
leading and non leading contributions in $g$ and, as such,
are not entirely consistent\cite{Heinz83}.
 One can also combine
 Eqs.~(\ref{4Spsi})
and (\ref{4kJgpsi}) into a single equation for  the total current density
induced by the fermionic fields,
  ${\cal J}^\psi\equiv {\cal J}^\psi_{\rm f}+{\cal J}^\psi_{\rm b},$
\beq
\label{4Jpsi}
\left [k\cdot D_X,\,{\cal J}^\psi_\mu(k,X) \right ]=ig\,k_\mu\,t^a
\,\left \{\bar\psi(X)\,t^a\,\slashchar{{\cal K}}(k,X)
-\slashchar{{\cal H}}(k,X)\,t^a\,\psi(X)\right \}.
\eeq
This equation is similar to the corresponding one in the abelian
case \cite{QED}. In doing  the sum of Eqs.~(\ref{4Spsi})
 and (\ref{4kJgpsi}), the typical non-abelian effects cancel; these are
contained for example in the second braces in Eq.~(\ref{4Spsi}), and involve
the
 3-gluon vertex leading to gauge field insertions on the  hard gluon lines.
 This kind of cancellation was first noted
by Taylor and Wong \cite{Taylor90} in relation with the HTL's for amplitudes
involving one pair of quarks and any number of soft gluons (albeit their
proof is only explicit up to three external gluons).

We have thus reduced the set of equations (\ref{4kinK}-\ref{covJg}) to
three fundamental equations, namely Eqs.~(\ref{4kinK}), (\ref{4JA}) and
(\ref{4Jpsi}) for the densities of the induced sources. From the
gauge-covariant character of these equations, it follows that, under local
gauge
 transformations,
$\eta^{ind}$ transforms like $\psi$, while $j^{ind}$
transforms like $F_{\mu\nu}$. Provided the external sources are chosen so
as to satisfy the same property, the mean fields equations
(\ref{4avpsi}) and (\ref{4avA}) are then gauge  covariant, as are the classical
equations of motion derived from the QCD action.

Eqs.~(\ref{4kinK},\ref{4JA},\ref{4Jpsi}) contain all the information on the
generalized polarizability of the plasma.
After
solving Eq.~(\ref{4kinK}), we compute the fermionic induced source
according to Eq.~(\ref{4eind}) and obtain
\beq
\label{eta-psi} \eta^{ind}(X)&=&
 -i\omega_0^2\int\frac{d\Omega}{4\pi}\slashchar{v}
\int_0^\infty duU(X,X-vu)\psi(X-vu)\nonumber\\
&\equiv&  \int d^4Y\delta \Sigma_A(X,Y) \psi(Y),
\eeq
with $\omega_0^2\equiv C_f(g^2 T^2/8)$ and $v^\mu\equiv (1,\,\vec v)$,
$\vec v\equiv \vec k/\epsilon_k$. The angular integral runs over all directions
of $\vec v$, and $U(x,y)$ is the parallel transporter along a straight line
joining
$x$ and $y$\cite{QED}. The kernel $\delta \Sigma_A$ is the self energy of a
soft fermion  propagating in  a background gauge field.
 The current induced by  fermionic mean fields results from Eq.~(\ref{4Jpsi}):
\beq
\label{4jpsi}
j_\mu^\psi(X)&=&g\omega_0^2 t^a\int\frac{d\Omega}{4\pi}v_\mu v_\nu\int_0^\infty
dt\int_0^\infty ds \nonumber\\
&&\qquad\bar\psi(X-vt)\gamma^\nu
U(X-vt,X) \, t^a\, U(X,X-vs)\psi(X-vs)\nonumber\\
\qquad &\equiv& g\, t^a\int d^4Y_1 d^4Y_2
\bar\psi(Y_1)\delta\Gamma_{\mu,\,a}^A(X;Y_1,Y_2)\psi(Y_2).
\eeq
The correction $\delta\Gamma^A$ to the
 quark-gluon vertex may be easily read out from this equation.
Finally, the current induced by soft gauge fields is determined from
Eq.~(\ref{4JA}) to be \beq
\label{4jA}
j^{A\, }_\mu(X)&=&3\,\omega^2_p\int\frac{d\Omega}{4\pi}
v_\mu\int_0^\infty du\, U(X,X-vu)\,F_{0\,j}(X-vu)v^j\,U(X-vu,X).
\eeq
Here $\omega^2_p\equiv (g^2 T^2/9)(N+N_f/2)$ is the plasma frequency.
By successive functional differentiation with respect to $A$ of
 Eq.~(\ref{4jA})
we derive corrections to the equilibrum  amplitudes for soft gluon fields.
All the amplitudes obtained in this way,
as well as those contained in $\delta\Sigma_A$ and in $\delta\Gamma_\mu^A$,
coincide with the HTL's of the diagrammatic approach.

\noindent
 Eqs.~(\ref{eta-psi}-\ref{4jA}) imply the following
covariant conservation laws for the induced currents\cite{QCD}:
\beq
\left[D^\mu,\,j_\mu^A\right]=0,
\eeq
and
\beq
\left[D^\mu,\,j_\mu^\psi\right]=igt^a\left(\bar\psi t^a\eta^{ind}-
\bar\eta^{ind}t^a\psi\right).
\eeq
 By differentiating these equations with respect to the
fields, one obtains relations between HTL refered as  ``QED-like Ward
identities''  in Ref.\cite{Frenkel90,Braaten90}.
Finally, by using these relations, together with the Jacobi identity $\left [
D^\mu,
 [ D^ \nu, F_{\nu\mu}]\right ]=0$, we see that the external sources must
satisfy
\beq
\left[D^\mu,\,j_\mu\right]=igt^a\left(\bar\psi t^a\eta-
\bar\eta t^a\psi\right).
\eeq
in order for the mean fields equations (\ref{4avpsi},\ref{4avA}) to be
consistent.

By eliminating the induced sources from Eqs.~(\ref{4avpsi},\ref{4avA}),
using their
explicit expressions (\ref{eta-psi}-\ref{4jA}), one obtains nonlinear
equations of
motion which generalize the Maxwell equations in a polarizable medium.
In particular, for vanishing external sources, these equations describe the
normal modes of the plasma. Note that in
general, as a consequence of gauge covariance,  quark and gluon modes mix.

These non linear equations for the mean fields can be generated by the minimal
 action principle applied to an effective action  $S_{eff}=S_0+S_{ind}$.
Here $S_0$ is the classical QCD action, while $S_{ind}$ contains the effects
of the interactions between
the soft fields and the hard particles of the plasma. It follows that $S_{ind}$
must satify $
\delta S_{ind}/\delta \bar\psi(X)=\eta^{ind}(X)$ and $
\delta S_{ind}/\delta A^\mu_a(X)=j^{ind\, a}_\mu(X)$.
These conditions are satisfied by $S_{ind}=S_f+S_g$, with \cite{QCD}
\beq
\label{Sf}
S_f&=& -i\omega_0^2\int\frac{d\Omega}{4\pi}\int
d^4X\int_0^\infty du\bar\psi(X)\slashchar{v}U(X,X-vu)\psi(X-vu),
\eeq and \beq
\label{Sg}
\qquad S_g&=&\frac{3}{2}\,\omega^2_p\int \frac{d\Omega}{4\pi}\int d^4 X
\int_0^\infty du\int_u^\infty du^\prime\nonumber\\
&&\qquad\,{\rm tr} \left\{v^\mu F_{\mu\lambda}(X)\,U(X,X-vu^\prime)\,
v_\nu F^{\nu\lambda}(X-vu^\prime)\,U(X-vu^\prime, X)\right \},
\eeq
where the trace acts on color indices only. This  gauge
invariant action coincides with the generating functional for HTL's derived
in \cite{Taylor90} on the basis of gauge invariance.
Here $S_{eff}$ has a different, more physical, interpretation: it is the
classical action describing long wavelength excitations in the hot quark-gluon
plasma, at leading order in the coupling $g$.

\end{document}